\documentclass[bm,aps,showpacs,amsfonts,amssymb,preprint,nofootinbib]{revtex4}

\usepackage{graphicx}
\usepackage{natbib}
\usepackage{amsmath}

\usepackage{color}

\font\mybb=msbm10 at 12pt

\def\bb#1{\hbox{\mybb#1}}

\def\ZZ {\bb{Z}}

\def\CCC {\bb{C}}
\def\PP {\bb{P}}

\def\FF {\bb{F}}

\def\nt{\tilde{n}}

\newcommand\beqa{\begin{eqnarray}}
\newcommand\eeqa{\end{eqnarray}}
\newcommand\n{\nonumber\\}

\begin{document}

{~}

\title{Note on three-generation models in heterotic string and 
F-theory on elliptic Calabi-Yau manifolds over Hirzebruch varieties}

\vspace{2cm}
\author{Shun'ya Mizoguchi\footnote[1]{E-mail:mizoguch@post.kek.jp} and Tomoki Sakaguchi\footnote[2]{E-mail:stomoki@post.kek.jp}}

\vspace{2cm}

\affiliation{\footnotemark[1]Theory Center, Institute of Particle and Nuclear Studies,
KEK\\Tsukuba, Ibaraki, 305-0801, Japan 
}
\affiliation{\footnotemark[1]\footnotemark[2]SOKENDAI (The Graduate University for Advanced Studies)\\
Tsukuba, Ibaraki, 305-0801, Japan 
}

\begin{abstract} 
We give a complete list of a class of  
three-generation models in $E_8\times E_8$ heterotic string theory 
and its dual F-theory on an elliptic Calabi-Yau over a (generalized) 
Hirzebruch variety in which the divisors of the relevant line bundles 
needed for a smooth Weierstrass model are effective. 
The most stringent constraint on the bound 
of the eta class comes from the effectiveness of the divisor 
of the bundle corresponding to the highest Casimir in Looijenga's 
weighted projective space, as well as from the compactness of the 
toric variety. Comparison is also made with the list 
obtained in the literature.

\end{abstract}

\preprint{KEK-TH 1910}
\date{July 7, 2016}

\maketitle

\newpage
The origin of three generations of flavors in particle physics is an enigma.
There has been no evidence found for the fourth new generation 
at the LHC experiment (see e.g.\cite{Vysotsky:2013gfa}), 
and the results of PLANCK has shown that 
the CMB data is consistent with the effective number 
of neutrinos derived by assuming three neutrino flavors 
(see e.g.\cite{Lesgourgues:2014zoa}).

Although so far none of the string compactifications can explain why 
there are three generations in nature, F-theory/$E_8\times E_8$ heterotic 
models have the following advantages over other  
string  (in particular D-brane)  models:  
(1) They naturally lead to $SU(5)$ GUT models which beautifully explain 
the observed hypercharge assignments, which are 
otherwise difficult to explain. (2) They can realize the spinor representation 
of $SO(10)$, which can contain all the quarks and leptons in one generation 
as a complete multiplet. (3) They can give up-type Yukawa couplings, which 
would be perturbatively forbidden in D-brane models.
(For a rather recent accumulation of literature on F-theory models, see 
the seminal papers \cite{DonagiWijnholt,BHV,BHV2,DonagiWijnholt2} 
and their citations.)

In the standard heterotic string compactification to four dimensions 
with a hermitian vector bundle $V$  \cite{FMW}, 
the number of chiral generation is given by, 
assuming $c_1(V)=0$, a half the third Chern class $\frac12 c_3(V)$.  
Some time ago,  
%
%
it was found \cite{Curio} that this was given, 
for a vector 
bundle characterized by the $\eta$ class (the first Chern class of the 
twisting line bundle, corresponding to the number of instantons) and 
the $\gamma$ class (the kernel of the projection from the spectral 
cover to the base which leads to some ambiguity of the vector bundle, 
corresponding to the $G$-flux in F-theory), as
\beqa
\frac12 c_3(V)&=&\lambda\eta(\eta-n c_1)
\label{thirdChern}
\eeqa
for an $SU(n)$ bundle, 
where $\lambda$ is a half-integral number related to the $\gamma$ class, 
and $c_1$ is the first Chern class of the base.

In \cite{Curio}, a list of $\eta$ classes leading to precisely three generations 
was also given, for $n=5,4$ and $3$ corresponding to the  $SU(5)$,  
$SO(10)$ and $E_6$ gauge group, respectively, for the base 
being a Hirzebruch surface or a del Pezzo surface. (See also \cite{Rajesh}, 
where the lower bound of the $\eta$ class was also mentioned.) 
Not all the divisors 
in the list, however,  
turn out to be not usable, since the relevant 
divisors are not effective and holomorphic sections do not exist.

In this letter, we will solve the equation (\ref{thirdChern}) 
in an elementary way for the case of the base being a Hirzebruch surface, 
for which the dual F-theory Calabi-Yau becomes an elliptic 
fibration over a generalized Hirzebruch variety\footnote{See \cite{Klemm:1996ts,Mohri,Rajesh} 
for earlier works on F-theory on elliptic fourfolds over toric varieties.},
to give 
a {\em complete} list of three-generation models with relevant effective divisors 
in this class of compactifications.

Of course, three-generation models in string theory are not rare,
\footnote{The literature on the construction of three-generation models 
is vast; a few of the notable papers include \cite{DHOR1,DHOR2,BD}.} 
nor does our result explain the mystery of the number of generations 
of quarks and leptons. We are, however, interested in these models because  
they are simple, and still infinitely many, so that they may serve as useful
models to 
study some fundamental questions in F-theory model buildings, such as 
the multiple singularity enhancement and associated family unification 
\cite{FFamilyUnification,MizoguchiTaniAnomaly}, computations of 
 Yukawa couplings and nonperturbative superpotentials, 
 mechanisms of moduli stabilization and supersymmetry breaking, etc.  
 We plan to explore these issues elsewhere.

The Hirzebruch surface $\FF_m$ $(m=0,1,2,\ldots)$ is a complex-dimension two manifold 
with coordinates $(z',w',z,w)\in\CCC^4$ with the identifications of points 
\beqa
(z',w',z,w)&\sim&(\mu z',\mu w',\kappa \mu^m z,\kappa w)
\eeqa
for arbitrary 
$\kappa,\mu\in\CCC^*=\CCC-\{0\}$,
where the points satisfying 
$z=w=0$ or $z'=w'=0$ are removed. The weights of the identifications 
are summarized by the table
\beqa
\begin{array}{ccccc}
&z'&w'&z&w\\
\mu~~~&1&1&m&0\\
\kappa~~~&0&0&1&1\\
&D_1&D_2&D_3&D_4
\end{array},
\label{Fmweight}
\eeqa
where we have also displayed the corresponding divisors 
in the bottom row. The associated toric fan is taken to be 
\beqa
\begin{array}{ccccc}
\phantom{\mu}~~~&1&-1&0&0\\
\phantom{\kappa}~~~&0&-m&1&-1\\
&\phantom{D_1}&\phantom{D_2}&\phantom{D_3}&\phantom{}
\end{array},
\label{Fmfan}
\eeqa
which is nothing but the orthogonal complement of (\ref{Fmweight}) 
if the numbers are viewed as a collection of row vectors.
(\ref{Fmfan}) implies the relations
\beqa
D_1&=&D_2,\n
D_3&=&D_4+m D_2.
\eeqa
Writing $D_1=D_2=f$, $D_4=C_0$ and $D_3=C_\infty$, 
they are known to have intersections 
\beqa
C_0^2=-m,~~~f^2=0,~~~C_0\cdot f=1.
\eeqa
The anticanonical class is
\beqa
c_1(\FF_m)&=&
2C_0+(2+m)f.
\eeqa

The generalized Hirzebruch variety $\FF_{mqp}$ 
$(m,q,p=0,1,2,\ldots)$ 
we consider is a complex-dimension 
three manifold with coordinates $(z'',w'',z',w',z,w)\in\CCC^6$, which are 
similarly 
subject to the identifications 
\beqa
(z'',w'',z',w',z,w)&\sim&(\nu z'', \nu w'',  \mu \nu^m z',\mu w',\kappa \mu^q \nu^p z,\kappa w)
\eeqa
for arbitrary $\kappa,\mu,\nu\in\CCC^*$,   
with the points satisfying 
$z=w=0$ or $z'=w'=0$ or $z''=w''=0$ being all removed. 
The weights and divisors are 
\beqa
\begin{array}{ccccccc}
&z''&w''&z'&w'&z&w\\
\nu~~~&1&1&m&0&p&0\\
\mu~~~&0&0&1&1&q&0\\
\kappa~~~&0&0&0&0&1&1\\
&D_1&D_2&D_3&D_4&D_5&D_6
\end{array}.
\label{Fmnpweight}
\eeqa
The corresponding fan 
\beqa
\begin{array}{ccccccc}
\phantom{\nu}~~~&1&-1&0&0&0&0\\
\phantom{\mu}~~~&0&-m&1&-1&0&0\\
\phantom{\kappa}~~~&0&-p&0&-q&1&-1\\
&&\phantom{D_2}&\phantom{D_3}&\phantom{}&\phantom{D}&\phantom{}
\end{array}
\label{Fmnpfan}
\eeqa
also implies the relations
\beqa
D_1&=&D_2,\n
D_3&=&D_4+m D_2,\n
D_5&=&D_6+p D_2 + q D_4,
\eeqa
so that the anticanonical class is ($D_2=f$, $D_4=C_0$)
\beqa
c_1(\FF_{mqp})&=&
(2+q)C_0+(2+m+p)f+2 D_6\n
&=&c_1(\FF_{m})+(q C_0 + p f)+2 D_6.
\label{Fmnpc1}
\eeqa
The ``base part'' of  (\ref{Fmnpc1}) differs from $c_1(\FF_{m})$ 
by $q C_0 + p f$ so that an elliptic fibration over $\FF_{mqp}$ 
can be used \cite{Rajesh} 
for a dual description of the heterotic compactification on 
the elliptic fibration over $\FF_m$ whose the vector bundle 
is (partly) characterized by the class  
$\eta=6c_1(\FF_{m})+ q C_0 + p f$ of the twisting line bundle \cite{FMW}.

Let us now solve (\ref{thirdChern}).
Putting 
\beqa
\eta&=&xC_0+yf~~~(x,y\in\ZZ),
\label{etaxy}
\eeqa
(\ref{thirdChern}) is equivalent to the equation
%
\beqa
\left(
\begin{array}{cc}
 x ~~&   y
 \end{array}
\right)
\left(
\begin{array}{cc}
 -m & 1  \\
  1  & 0  
\end{array}
\right)
\left(
\begin{array}{c}
x-2n  \\
y-(2+m)n  
\end{array}
\right)&=&\frac3\lambda,
\eeqa
which can be solved for $y$ as
\beqa
y&=&\frac m2 x+n+\frac{n^2+\frac3{2\lambda}}{x-n}.
\label{ym2x}
\eeqa
$\lambda$ is taken to be $\pm\frac12$ or $\pm\frac32$, and only 
for $n=4$, $\lambda=\pm 1, \pm 3$ is also allowed. 
Since $y$ is an integer, the right hand side of (\ref{ym2x}) must also
 sum up to an integer. If $\lambda\in\ZZ+\frac12$, This 
implies the following:
\begin{itemize}
\item[({\bf I})]  If $mx\in 2\ZZ$, then 
the integer $n^2+\frac3{2\lambda}$ must be divisible by $x-n$. 

\item[({\bf II})] If  $mx\in 2\ZZ+1$, then 
the integer $2(n^2+\frac3{2\lambda})$ must be divisible by $x-n$, 
and at the same time the quotient must be an odd integer. 
\end{itemize}

In addition, if $n=4$ and $\lambda=\pm1$ or $\pm 3$, then 
\begin{itemize}
\item[({\bf I'})] 
If $mx\in 2\ZZ$, then 
the integer $2(n^2+\frac3{2\lambda})$ must be divisible by $x-n$, 
and at the same time the quotient must be an even integer. 

\item[({\bf II'})] 
If $mx\in 2\ZZ+1$, then 
the integer $2(n^2+\frac3{2\lambda})$ must be divisible by $x-n$, 
and at the same time the quotient must be an odd integer. 
\end{itemize}

Not all the solutions to the above are suitable for smooth compactifications. 
First of all, both $\eta$ and 
$12c_1 - \eta$ must be effective (= all the coefficients of the 
divisor are non-negative), because they are the classes of the 
twisting line bundles of the spectral cover. 
The physical interpretation of this condition 
is that the instanton number of one of $E_8$ does not exceed 24 
for each $K3$ fiber in the heterotic string compactification.  
In fact, the consideration of Looijenga's weighted projective space bundle 
\cite{FMW,MizoguchiTaniLooijenga}
provides a more stringent constraint on the possible $\eta$. 
Looijenga's weighted projective space is known as the 
moduli space of the vector bundle for the heterotic compactifications, 
and the section of the bundle consists of the independent 
polynomials in the coefficients of the Weierstrass model in F-theory. 
In the $SU(5)$ case, they are the polynomials 
\beqa
h_{\nt+2},~H_{\nt+4},~p_{\nt+6},f_{\nt+8}~\mbox{and}~g_{\nt+12}
\eeqa 
appearing 
in the well-known {\em six}-dimensional F-theory 
\cite{MorrisonVafa,BIKMSV} 
compactified on an elliptic CY3 over a Hirzebruch surface 
$\FF_{\nt}$, where the subscripts denote the degrees of the polynomials. 
More generally (in four dimensions), they are the sections of the line bundles 
\beqa
a_{1,0}&\in&\Gamma({\cal L}^{-5}\otimes {\cal N}),\n
a_{2,1}&\in&\Gamma({\cal L}^{-4}\otimes {\cal N}),\n
a_{3,2}&\in&\Gamma({\cal L}^{-3}\otimes {\cal N}),\n
a_{4,3}&\in&\Gamma({\cal L}^{-2}\otimes {\cal N}),\n
a_{6,5}&\in&\Gamma({\cal L}^{-0}\otimes {\cal N})
\eeqa
\cite{FMW,DonagiWijnholt,MizoguchiTaniAnomaly,MizoguchiTaniLooijenga}, 
where ${\cal L}$ is the anticanonical line bundle of the base of 
the heterotic threefold, and ${\cal N}$ is the twisting line bundle for 
the vector bundle over this threefold. 
$\Gamma$ denotes the space of sections. The divisors 
of these bundles must be effective. The most stringent constraint 
comes from $a_{1,0}$ ($\sim h_{\nt+2}$), which asserts that $\eta-5c_1$ 
must be effective. Thus we have
\beqa
10\leq x\leq 24,~~~10+5m \leq y \leq 24 + 12m.
\eeqa

For $SO(10)$, the section $a_{1,0}$ ($\sim h_{\nt+2}$) becomes zero, 
but instead $a_{2,1}$ ($\sim H_{\nt+4}$) must exist. Therefore $\eta-4c_1$ 
must be effective.  Similarly, $a_{2,1}$ becomes zero for $E_6$, 
and then for $a_{3,2}$ ($\sim p_{\nt+6}$) to exist $\eta-3c_1$ must be effective. 
Thus, in general 
\beqa
2n\leq x\leq 24,~~~(2+m)n \leq y \leq 24 + 12m \label{xyrange}
\eeqa
for $n=5,4,3$.

Before displaying the list of the solutions, 
a remark is in order about the compactness of the generalized Hirzebruch 
variety defined by the weights (\ref{Fmnpweight}): 
We can assume $m\geq 0$ without loss of generality since, if $m<0$, 
we can consider the new scaling $\nu\mu^{-m}$ and 
interchange the roles of $z'$ and $w'$. 
Similarly, if both of $(q,p)=(x-12,y-6m-12)$ happen to be 
negative, we can consider the scalings $\nu\kappa^{-p}$ and $\mu\kappa^{-q}$ 
replace $z$ with $w$ and vice versa; this corresponds to consider 
the unbroken gauge group residing ``at infinity'' of the $\PP^1$ fiber. 
If, however, only one of $(q,p)$ happens to be negative, 
there is no way to define such a new scaling and the variety 
would become noncompact. Therefore we restrict ourselves to 
the solutions such that {\em both} of $(q,p)=(x-12,y-6m-12)$ are 
non-negative, or {\em both} of $(q,p)=(x-12,y-6m-12)$ are non-positive.

The complete list of the solutions satisfying the condition 
({\bf I}) or ({\bf II}) or ({\bf I'}) or ({\bf II'}) 
in the range (\ref{xyrange}), and satisfying the condition 
in the remark above,  is as follows:

\noindent
\underline{$SU(5)$ models ($n=5$)}
\begin{itemize}
\item{$\lambda=\frac12$}
\beqa
(x,y)&=&(12,6m+9)~~~(m\in\ZZ,~m\geq1),\n
&&(19,\frac{19}2 m+7)~~~(m\in2\ZZ,~m\geq2),\n
&&(13,\frac{13}2 m+\frac{17}2)~~~(m\in2\ZZ+1,~m\geq7
).\nonumber
\eeqa
\item{$\lambda=-\frac12$}
\beqa
(x,y)&=&(16,8m+7)~~~(m\in\ZZ,~m\geq3
).\nonumber
\eeqa
\item{$\lambda=\frac32$}
\beqa
(x,y)&=&(18,9m+7)~~~(m\in\ZZ,~m\geq2
).\nonumber
\eeqa
\item{$\lambda=-\frac32$}
\beqa
(x,y)&=&(11,\frac{11}2m+9)~~~(m\in2\ZZ,~m\geq2),\n
&&(13,\frac{13}2 m+8)~~~(m\in2\ZZ,~m\geq8
),\n
&&(17,\frac{17}2 m+7)~~~(m\in2\ZZ,~m\geq2),\n
&&(21,\frac{21}2 m+\frac{13}2)~~~(m\in2\ZZ+1,~m\geq3
).\nonumber
\eeqa
\end{itemize}

\noindent
\underline{$SO(10)$ models ($n=4$)}
\begin{itemize}
\item{$\lambda=\frac12$}
\beqa
(x,y)&=&(23,\frac{23}2 m+5)~~~(m\in2\ZZ,~m\geq2).\nonumber
\eeqa
\item{$\lambda=-\frac12$}
\beqa
(x,y)&=&(17,\frac{17}2 m+5)~~~(m\in2\ZZ,~m\geq4
).\nonumber
\eeqa
\item{$\lambda=\frac32$}
\beqa
(x,y)&=&(21,\frac{21}2 m+5)~~~(m\in2\ZZ,~m\geq2).\nonumber
\eeqa
\item{$\lambda=-\frac32$}
\beqa
(x,y)&=&(9,\frac{9}2 m+7)~~~(m\in2\ZZ,~m\geq2),\n
&&(19,\frac{19}2 m+5)~~~(m\in2\ZZ,~m\geq2).\nonumber
\eeqa
\item{$\lambda=1$}
\beqa
(x,y)&=&(9,\frac{9}2 m+\frac{15}2)~~~(m\in2\ZZ+1,~m\geq1),\n
&&(11,\frac{11}2 m+\frac{13}2)~~~(m\in2\ZZ+1,~m\geq1),\nonumber
\eeqa
\item{$\lambda=3$}
\beqa
(x,y)&=&(15,\frac{15}2 m+\frac{11}2)~~~(m\in2\ZZ+1,~m\geq5
),\nonumber
\eeqa
\item{$\lambda=-1,-3$}\\~~~~~~~~~~~~~~~~~~~~~~~
No solution.
\end{itemize}

\noindent
\underline{$E_6$ models ($n=3$)}
\begin{itemize}
\item{$\lambda=\frac12$}
\beqa
(x,y)&=&(6,3 m+7)~~~(m\in\ZZ,~m\geq0),\n
&&(7,\frac{7}2 m+6)~~~(m\in2\ZZ,~m\geq0),\n
&&(9,\frac{9}2 m+5)~~~(m\in2\ZZ,~m\geq2),\n
&&(15,\frac{15}2 m+4)~~~(m\in2\ZZ,~m\geq6
),\n
&&(11,\frac{11}2 m+\frac92)~~~(m\in2\ZZ+1,~m\geq1).\nonumber
\eeqa
\item{$\lambda=-\frac12$}
\beqa
(x,y)&=&(9,\frac{9}2 m+4)~~~(m\in2\ZZ,~m\geq2),\n
&&(7,\frac{7}2 m+\frac92)~~~(m\in2\ZZ+1,~m\geq3),\n
&&(15,\frac{15}2 m+\frac72)~~~(m\in2\ZZ+1,~m\geq7
).\nonumber
\eeqa
\item{$\lambda=\frac32$}
\beqa
(x,y)&=&(8,4 m+5)~~~(m\in\ZZ,~m\geq1),\n
&&(13,\frac{13}2 m+4)~~~(m\in2\ZZ,~m\geq16
),\n
&&(7,\frac{7}2 m+\frac{11}2)~~~(m\in2\ZZ+1,~m\geq1),\n
&&(23,\frac{23}2 m+\frac72)~~~(m\in2\ZZ+1,~m\geq3
).\nonumber
\eeqa
\item{$\lambda=-\frac32$}
\beqa
(x,y)&=&(7,\frac{7}2 m+5)~~~(m\in2\ZZ,~m\geq2),\n
&&(11,\frac{11}2 m+4)~~~(m\in2\ZZ,~m\geq2),\n
&&(19,\frac{19}2 m+\frac{7}2)~~~(m\in2\ZZ+1,~m\geq3
).\nonumber
\eeqa
\end{itemize}
One can verify that all the solutions listed above give 
precisely three generations while satisfying the condition (\ref{xyrange}). 
This is the main result of this letter.

In most cases, $m=0$ is not allowed. The rare exceptions are 
the first two solutions for $E_6$ with $\lambda=\frac12$.

We should note that
the equation (\ref{thirdChern}) has an obvious symmetry 
$\eta\rightarrow \tilde\eta=nc_1-\eta$, so if some $\eta$ satisfies (\ref{thirdChern}),
so does $\tilde\eta$, but this $\tilde\eta$ may or may not be 
a suitable divisor for which all the relevant divisors are effective.
The $\FF_{2k+1}$ solution $\eta=(-3,-3k)$ for $SU(5)$ in ref.\cite{Curio} 
has  $\tilde\eta=(13,13k+15)$, which coincides with our solution 
$(13,\frac{13}2 m+\frac{17}2)$ with $m=2k+1$. 
Unlike our solution, however, neither $(-3,-3k)$ nor $(-3,-3k)-5c_1$ 
is effective. The $\FF_{2k}$ solution $\eta=(3,3(k-2))$ for $SU(5)$ in 
ref.\cite{Curio}, or 
its $\tilde\eta$, is not contained in the above list because 
neither $\eta-5c_1$ nor $\tilde\eta-5c_1$ is effective. 
Also, the $SO(10)$ or $E_6$ solutions found in \cite{Curio} 
or their $\tilde\eta$ do not coincide with any of our solutions since 
their relevant divisors are not effective, either. The ``closest'' is 
$\eta=(0,1)$ for $E_6$ ($\lambda=-\frac12$) which has $\tilde\eta=(6,3m+5)$ 
with a marginal value for $x$ $(=6)$, but $y=3m+5$ is outside 
the range (\ref{xyrange}).

To summarize, we have shown a complete list of $\eta$ classes leading to 
precisely three-generations in $E_8\times E_8$ heterotic string theory 
on an elliptic Calabi-Yau over a Hirzebruch surface and its dual F-theory 
on an elliptic Calabi-Yau over a generalized Hirzebruch variety, 
where the divisors of the relevant line bundles 
needed for a smooth Weierstrass model are all effective. 
We hope they will be used as 
a concrete simple model to investigate fundamental questions in F-theory 
as we mentioned in the beginning of this letter.

We thank K.~Kohri, K.~Mohri and T.~Tani for discussions. 
The work of S.~M. is supported by 
Grant-in-Aid
for Scientific Research  
(C) \#25400285, 
(C) \#16K05337 
and 
(A) \#26247042
from
The Ministry of Education, Culture, Sports, Science
and Technology of Japan.

\end{document}